\documentclass[prd, aps, reprint, amsfonts, amssymb, preprintnumbers, showpacs, nofootinbib, superscriptaddress]{revtex4-1}
\usepackage{graphicx, multirow,soul}
\usepackage[citecolor=blue]{hyperref}
\usepackage[all]{hypcap}
\usepackage{url}
\usepackage{color}
\usepackage{subfigure}
\usepackage{slashed}
\usepackage{bbding}
\usepackage{amsmath}
\usepackage{pifont}
\newcommand{\cmark}{\text{\ding{51}}}
\newcommand{\xmark}{\text{\ding{55}}}
\usepackage{wasysym}
\usepackage{amssymb}
\usepackage{float}
\usepackage{bbm}
\usepackage{bbold}
\usepackage{lipsum}
\newcommand{\equaref}[1]{Eq.~(\ref{#1})}

\newcommand{\figref}[1]{Fig.~\ref{#1}}

\newcommand{\secref}[1]{Section~\ref{#1}}

\newcommand{\tabref}[1]{Table~\ref{#1}}

\usepackage{tikz}
\tikzset{node distance=2cm, auto}
\usepackage{bm}
\newcommand{\bq}{\begin{eqnarray}}
\newcommand{\nq}{\end{eqnarray}}

\definecolor{myblue}{rgb}{0.39, 0.589, 0.6914}
\newcommand{\eV}{{\rm eV}}

\newcommand{\GeV}{{\rm GeV}}

\begin{document}

\title{Tau neutrinos at DUNE: new strategies, new opportunities}

\author{Pedro Machado}
\affiliation{Theoretical Physics Department, Fermi National Accelerator Laboratory, P.O. Box 500, Batavia, IL 60510, USA.}
\author{Holger Schulz}
\affiliation{Department of Physics, University of Cincinnati, Cincinnati, OH 45219, USA.}
\author{Jessica  Turner}
\affiliation{Theoretical Physics Department, Fermi National Accelerator Laboratory, P.O. Box 500, Batavia, IL 60510, USA.}

\date{\today}

\begin{abstract}
We propose a novel analysis strategy, that leverages the unique capabilities of the DUNE experiment,
 to study tau neutrinos. 
We integrate collider physics ideas, such as jet clustering algorithms in combination with machine learning
techniques,  into neutrino measurements. 
Through the construction of a set of observables and kinematic cuts, we obtain a
superior discrimination of the signal ($S$) over the background ($B$).
In a single year, using the nominal neutrino beam mode, 
DUNE may achieve $S/\sqrt{B}$ of $3.3$ and $2.3$ for
the hadronic  and leptonic decay channels of the  tau respectively.
Operating in the  tau-optimized beam mode would increase $S/\sqrt{B}$ to
$8.8$ and $11$ for each of these channels.
We premier the use of the analysis software
Rivet,  a tool ubiquitously used by the LHC experiments, in neutrino physics. 
For wider accessibility, we provide our analysis code.
\end{abstract}
\preprint{FERMILAB-PUB-20-257-T} 
\maketitle

\section{Introduction}
Arguably, the tau neutrino is the least understood particle of the Standard Model.
Thus far a total of 14 tau neutrinos have been positively identified by the DONuT~\cite{Kodama:2007aa} and 
OPERA~\cite{Agafonova:2018auq} experiments.
The former detected beam tau neutrinos from the decay of $D_s$ mesons. The latter observed, for the first time, 
$\nu_\mu\to \nu_\tau$ oscillations.
In both experiments, the identification of $\tau$ leptons produced by $\nu_\tau$ charged current (CC) interactions
 relies upon the reconstruction of characteristic event topologies: the $\tau$ lepton leaves a  millimeter-scale track in the detector emulsion followed by a kink from its subsequent decay.
In addition to  DONuT and OPERA, the presence of tau neutrinos has been  statistically  inferred from  $\nu_\mu\to\nu_\tau$
 oscillations of multi-GeV atmospheric neutrinos by Super-Kamiokande~\cite{Abe:2012jj, Li:2017dbe} and IceCube~\cite{Aartsen:2019tjl}.
These searches are based on the $\nu_\tau$ contribution to the number of hadronic and/or leptonic neutrino events.
 
 Despite the excellent  reconstruction capabilities  of  DONuT and OPERA and the large statistics of Super-Kamiokande and 
IceCube, tau neutrino observables, such as $\nu_\tau$-nuclei cross sections and oscillation parameters extracted 
from $\nu_\tau$ measurements, have large statistical and systematic uncertainties. 
Indeed, the tau neutrino nucleon interaction cross section has larger uncertainties \cite{Kodama:2007aa, Jeong:2010nt, Li:2017dbe, Aartsen:2019tjl} than its 
electronic \cite{Blietschau:1977mu, Wolcott:2015hda, Acciarri:2020lhp} and muonic \cite{Agashe:2014kda, Abratenko:2020acr, Filkins:2020xol} counterparts. 
Super-Kamiokande provided the most accurate measurement of the $\nu_\tau$ CC  cross section using atmospheric neutrinos
and this
has an uncertainty of 21\%~\cite{Li:2017dbe}. Moreover, the unoscillated 
tau neutrino flux itself is a source of systematic uncertainty as it depends on the 
$D_s$ production rate; the accuracy of which is limited by the incomplete understanding of hadronic
effects. These two sources of uncertainty can be mitigated 
via the direct study of tau neutrino production as proposed by the DsTau experiment  \cite{Aoki:2019jry}. This
collaboration aims to provide an independent $\nu_\tau$ flux prediction for future neutrino beams with an
uncertainty below 10$\%$. As such, the  systematic error of the CC $\nu_\tau$  cross section
prediction can be lowered. This measurement, together with the use of near-to-far detector ratios,
 will be crucial in limiting systematic uncertainties in future studies of tau neutrinos at long-baseline neutrino experiments.

The observation of tau neutrinos in  current neutrino beams relies on tau appearance due to oscillations. 
The phase of the oscillation is given by terms such as 
\begin{equation}\label{eq:oscillations}
  \sin^2\left(\frac{\Delta m^2 L}{4E}\right)\simeq\sin^2\left(1.27\frac{\Delta m^2/\eV^2 L/\rm{km}}{E_\nu/\GeV}\right),
\end{equation}
where $\Delta m^2$ is a mass squared splitting, $E_\nu$ is the neutrino energy and $L$ is the experimental baseline.
The larger of the two mass squared splittings is  $\Delta m^2_{\rm atm}\simeq2.5\times10^{-3}~\eV^2$ and the neutrino energy threshold to create a tau lepton from CC interactions is $E_\nu\gtrsim 3.5~\GeV$.
Therefore, the baseline necessary to maximize $\nu_\tau$ appearance is of the order $2000-3000$~km.
At such long distances, accelerator neutrino experiments require extremely powerful neutrino sources to amass sufficient statistics in order to study the $\nu_\tau$ sector. A  further difficulty associated with $\nu_\tau$ detection is that the
decay of a $\tau$ lepton always includes a $\nu_\tau$ in the final state which carries away a fraction of undetectable energy. Therefore, 
reconstruction of the original beam neutrino energy is a challenging task. 
Although 
atmospheric neutrinos  have energies well above the tau production threshold and travel the necessary distances to
 induce large $\nu_\mu\to\nu_\tau$ oscillations, reconstructing the  energy of these neutrinos is
 challenging as the direction of the incident particle is not known on an event-by-event basis.

Despite such difficulties, upcoming multi-purpose  neutrino experiments, such as DUNE~\cite{Abi:2020wmh}, are well posed to detect tau neutrinos
given their large fiducial volume, powerful neutrino beam and exquisite track reconstruction capability. 
There have been a number of works which explore the tau neutrino sector including pioneering proposals
for observing $\nu_\tau$ at beam dump experiments \cite{Albright:1978ni} and the subsequent experimental search by NOMAD \cite{Astier:2001yj}, polarization effects on $\tau$ decay products
 for atmospheric and beam $\nu_\tau$ searches~\cite{Hagiwara:2003di, Levy:2004rk, Hagiwara:2004gs, Aoki:2005wb, Sobczyk:2019urm},
  optimization of  cuts on $\tau$ decay products in the DUNE beam neutrino sample~\cite{Conrad:2010mh}, tests of unitarity and patterns of 
  leptonic mixing~\cite{Parke:2015goa, Denton:2020exu, Ellis:2020ehi} and 
 beyond the Standard Model physics probes with $\nu_\tau$ measurements~\cite{Rashed:2016rda, Meloni:2019pse, deGouvea:2019ozk, Kling:2020iar}.

In this paper, we are particularly interested in future liquid argon time projection chamber (LArTPC) experiments such as DUNE which have 
proven to be endowed with excellent event topology reconstruction capabilities.
The future DUNE experiment will combine bubble chamber quality data with calorimetry and 
large statistics. It will therefore provide an unprecedented opportunity to study the $\nu_\tau$ sector.
We begin in \secref{sec:tausatdune} with a general discussion of the DUNE experiment and  tau neutrino detection at LArTPCs.
We  perform a sophisticated simulation of the tau neutrino signal and background processes in DUNE, taking into account tau lepton polarization and nuclear physics effects.  This 
is outlined in \secref{sec:results} where we build on and expand the analysis performed in Ref.~\cite{Conrad:2010mh} by using  modern techniques, such as jet clustering algorithms 
and deep neural networks, in order to  optimize signal-to-background ratios. Furthermore, we 
quantify the importance of charge identification of pions and running in the tau-optimized beam configuration for tau neutrino
searches. Finally, we summarize and discuss future studies, which will incorporate detector effects, in \secref{sec:summary}.

\section{Tau neutrinos at DUNE}\label{sec:tausatdune}
Before discussing tau neutrino events in detail, we first provide a description of the key aspects of the DUNE experiment.
The neutrino beam at DUNE is produced by a 120 GeV proton beam hitting a target.  The nominal beam power will be approximately 1.2~MW
and is expected to deliver $1.1\times10^{21}$ protons on target (POT) per year. The far detector consists of a 40 kiloton fiducial mass
 LArTPC with a baseline of 1300 km. 

\begin{figure}[t!]
\centering
\includegraphics[width=0.48\textwidth]{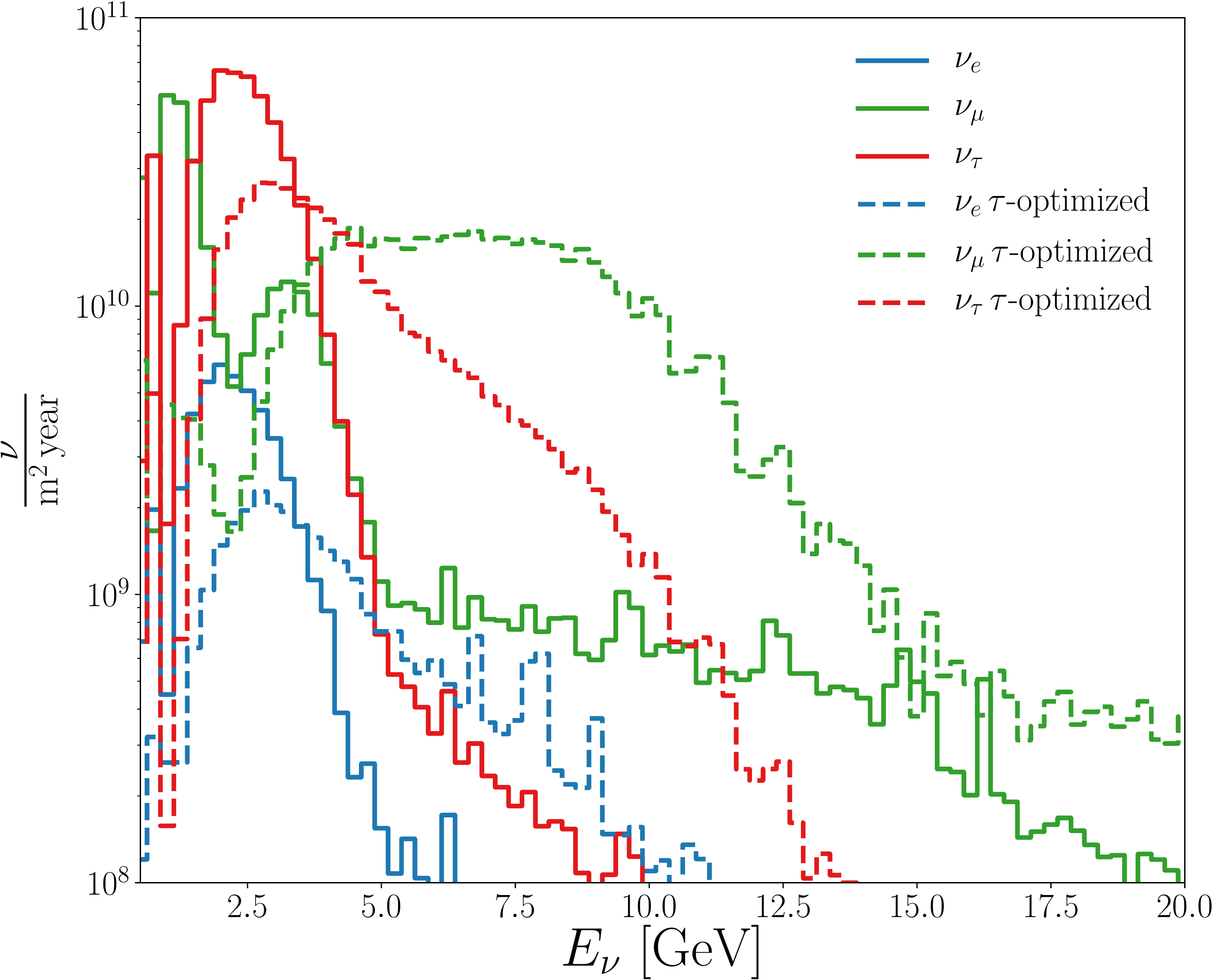}\label{fig:FDflux}
\caption{Nominal (solid) and tau-optimized (dashed) neutrino fluxes  at the DUNE far detector neutrino for $\nu_e$ (blue), $\nu_\mu$ (green) and $\nu_\tau$ (red). \label{fig:fluxes}}
\end{figure}

In \figref{fig:fluxes} we show the oscillated neutrino fluxes at the DUNE far detector for the nominal neutrino mode (solid) and tau-optimized 
configuration (dashed) for each neutrino flavor \cite{DUNEfluxes}.  
The oscillation parameters chosen as inputs throughout this paper are taken from global fit data~\cite{Esteban:2018azc}:
\begin{equation*}
\begin{aligned}
 \Delta m^2_{21} &= 7.4\times 10^{-5}~{\rm eV}^2, \quad  \Delta m^2_{31}= 2.5\times 10^{-3}~{\rm eV}^2, \\
 \quad s^2_{12} &= 0.31,\quad s^2_{13} = 0.0224, \quad
 s^2_{23}=0.5, \quad \delta_{CP}=1.2\pi,
 \end{aligned}
 \end{equation*}
  where $s_{ij}\equiv\sin\theta_{ij}$. In the nominal neutrino mode, the fluxes of all three neutrino flavors peak in the range $1$-$3$ GeV. 
Therefore many of the 
tau neutrinos (and antineutrino contaminants) have energies below the tau lepton production threshold.
While the integrated flux of the tau-optimized mode is similar to the nominal mode,  the peak of the  spectra for all three flavors is  broader 
and consequently there are more 
tau neutrinos with energies above the tau production threshold at the far detector.

In this study, we do not consider $\mu/\pi$ misidentification or energy and angular resolutions.
Furthermore, we  assume that particles above a certain  energy threshold will be identified and reconstructed with $100\%$ efficiency 
 in the detector~\cite{Abi:2020evt} as detailed in \secref{sec:results}.
We acknowledge our approach is optimistic as we do not apply detector effects. Moreover, application of such  effects will inevitably 
deteriorate the signal-to-background ratio; however, the purpose of this paper is to 
establish a new tool chain and  apply novel techniques which 
will be of use in a more detailed future study.

Conservatively, we assume neutrons are completely invisible at LArTPCs.
However, the techniques implemented in this work could benefit from information on the energy deposited by secondary hard neutron-proton scattering or multiple neutron scatterings. Such processes can constitute  a considerable fraction of the total hadronic energy of an event~\cite{Friedland:2018vry}.
Exploring this possibility would require a fully-fledged detector simulation which is beyond the scope of this paper. 

Finally, we exploit DUNE's capability to identify the charge of pions via the topology of 
the pion tracks (see e.g. Refs.~\cite{Nutini:2015lea, Gramellini:2018mjg, Abi:2020evt}).
Charged pions can undergo many different processes as they travel through a dense medium.
These processes include two-body pion absorption ($\pi^+np\to pp$, $\pi^-np\to nn$),  elastic scattering, single charge exchange  ($\pi^+n\to\pi^0p$, $\pi^-p\to\pi^0n$) and inelastic scattering. Most importantly, 
 stopped $\pi^-$ are typically captured by the positively charged argon nucleus as opposed to stopped $\pi^+$ which simply decay to $\mu^+$ followed by a Michel $e^+$. Due to these distinctively different topologies, charge identification of pions is possible on an event-by-event 
 basis. It should be noted there  have also been discussions on exploiting topological information~\cite{Suzuki:1987jf, Acciarri:2017sjy} to statistically distinguish the
 charge of muons, see e.g. Section 5.5.2.1 of Ref.~\cite{Abi:2020evt}.
In principle, similar techniques could be used to infer charge identification of pions statistically~\cite{Duffy:talk}.
We study the impact of perfect pion charge identification on signal and background yields and contrast this 
with the case $\pi^+$ cannot be distinguished from $\pi^-$.

Now that we have discussed the details of the DUNE experiment and the assumptions underlying our analysis, we proceed onto the physics of taus and tau neutrinos. 
As the DUNE beam is predominantly comprised of muon neutrinos, the main sample of tau neutrinos  at the far detector are due to $\nu_\mu\to\nu_\tau$ oscillations.
At the oscillation minimum of  $E_\nu \sim 3$ GeV, the majority of muon neutrinos are expected to have oscillated to tau neutrinos.

A key element in the study of tau neutrino physics is the decay modes of the tau lepton. The most relevant  tau branching ratios are given in
 \tabref{tab:branchings}. In the following, we denote the sample of taus that decay to electrons and muons as $\tau_e$ and $\tau_\mu$ respectively  while hadronically decaying taus will be denoted as $\tau_{\rm had}$.
The tau decay length has a value of $c\tau\approx87\,\mu$m, which is much larger than the argon nuclear radius (of about 3.4~fm), and 
thus the tau decays far outside the nucleus. Subsequently, its decay products are not subject to the argon's nuclear potential. 
 However, the tau lifetime is too short to lead to observable displaced vertices in DUNE where the granularity is limited by the typical 
 wire spacing of a few millimeters. It is thus unlikely that tau tracks can be observed at DUNE\footnote{It is possible a handful of tau tracks could be 
 observed from atmospheric tau neutrinos as they have very high energies.}. This, together with the severe background, makes
  $\nu_\tau$ detection particularly challenging. 
 
The background of the $\tau_\mu$ signal stems mainly from $\nu_\mu$ CC events\footnote{There is a subdominant contribution from the $\overline{\nu}_{\mu}$ CC interactions.}. This
channel is widely considered to be experimentally
 intractable as the $\nu_\mu$ flux  is prohibitively large.
Similarly, the dominant background of the $\tau_e$ signal are $\nu_e$ CC events.  As 
 the $\nu_e$ flux at DUNE is a small fraction of the total neutrino
 flux, we study $\nu_\tau$ detection in this channel in \secref{sec:leptonic}.

\begin{table}[t]
\begin{center}
\begin{tabular}{|c|c|} \hline
   \textbf{Decay mode}  &  \textbf{Branching ratio} \\ \hline \hline
   Leptonic & 35.2\% \\ \hline
   $e^-\bar\nu_e\nu_\tau$  &  17.8\% \\
   $\mu^-\bar\nu_\mu\nu_\tau$  &  17.4\% \\ \hline \hline
   Hadronic & 64.8\% \\ \hline
   $\pi^-\pi^0\nu_\tau$  &  25.5\% \\
   $\pi^-\nu_\tau$  &  10.8\% \\
   $\pi^-\pi^0\pi^0\nu_\tau$  &  9.3\% \\
   $\pi^-\pi^-\pi^+\nu_\tau$  &  9.0\% \\
   $\pi^-\pi^-\pi^+\pi^0\nu_\tau$  &  4.5\% \\
   other   &  5.7\% \\ \hline
\end{tabular}
\end{center}
\caption{Dominant decay modes of $\tau^-$. All decays involving kaons, as well as other subdominant decays, 
are in the ``other'' category.\label{tab:branchings}}
\end{table}

Finally, the dominant background to $\tau_{\rm had}$ are the neutral current (NC) neutrino scattering events which
have contributions from all three neutrino flavors.  Despite the fact that all neutrino flavors contribute to NC events (including tau neutrinos),
 the NC cross section is smaller than  the CC cross section \cite{Zeller:2001hh,Formaggio:2013kya}. Furthermore, the hadronic branching fraction of taus is almost
  twice as large as the leptonic branching fraction (see \tabref{tab:branchings}). Consequently, the hadronic decays of the tau have a higher signal-to-background ratio, in the nominal beam mode, than either of the leptonic channels
as we outline in \secref{sec:hadronic}. We note that there is a small contribution
  to the signal and background from $\overline{\nu}_{\tau}$ CC and $\overline{\nu}_{e,\mu,\tau}$ NC events respectively which we include in our analysis.  

\begin{figure}[t!]
\includegraphics[width=0.48\textwidth]{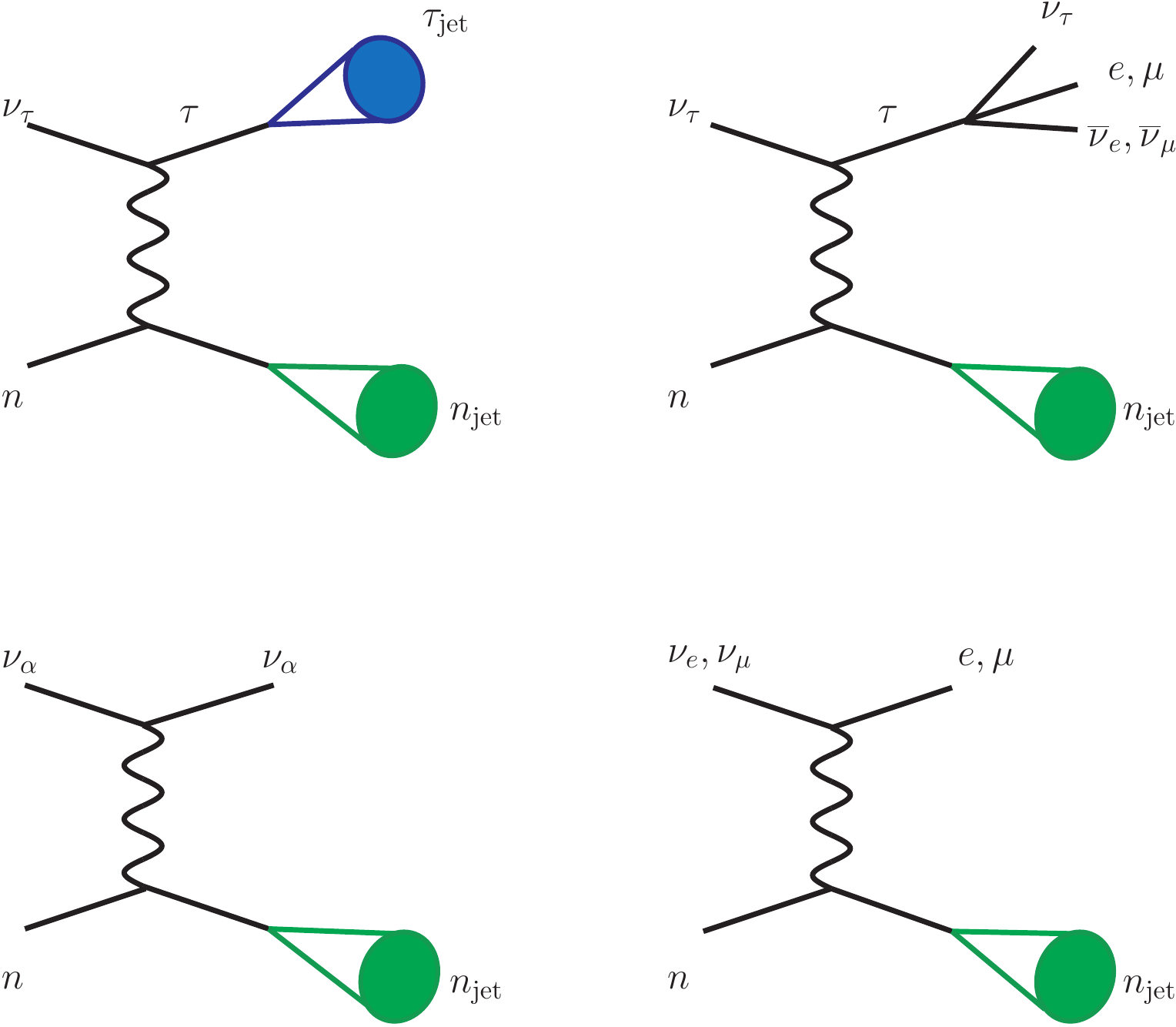}\label{fig:diagrams}
\caption{Pictorial representation of hadronic tau (upper left) and leptonic tau (upper right) signals, and their corresponding backgrounds (lower).}
\end{figure}

A pictorial summary of the dominant tau signals and backgrounds is shown in  \figref{fig:diagrams}.
In the upper right (left)  we show the leptonic (hadronic) decay of the tau and in the  lower 
right (left) its  associated dominant  background.
The target nucleon inside the argon nucleus is denoted as $n$ and the green cones represent the (mostly)
hadronic activity that emerges from the argon nucleus after the hadronization and subsequent intra-nuclear cascade. We denote this collection
of particles, emerging from the nucleus, as $n_{\rm jet}$ because we apply a jet clustering algorithm to the signal and background events.
Likewise, in the case of the hadronic decays of the tau we  represent the final states as $\tau_{\rm jet}$. 
%
\begin{figure*}[t!]
\centering
\includegraphics[width=0.95\textwidth]{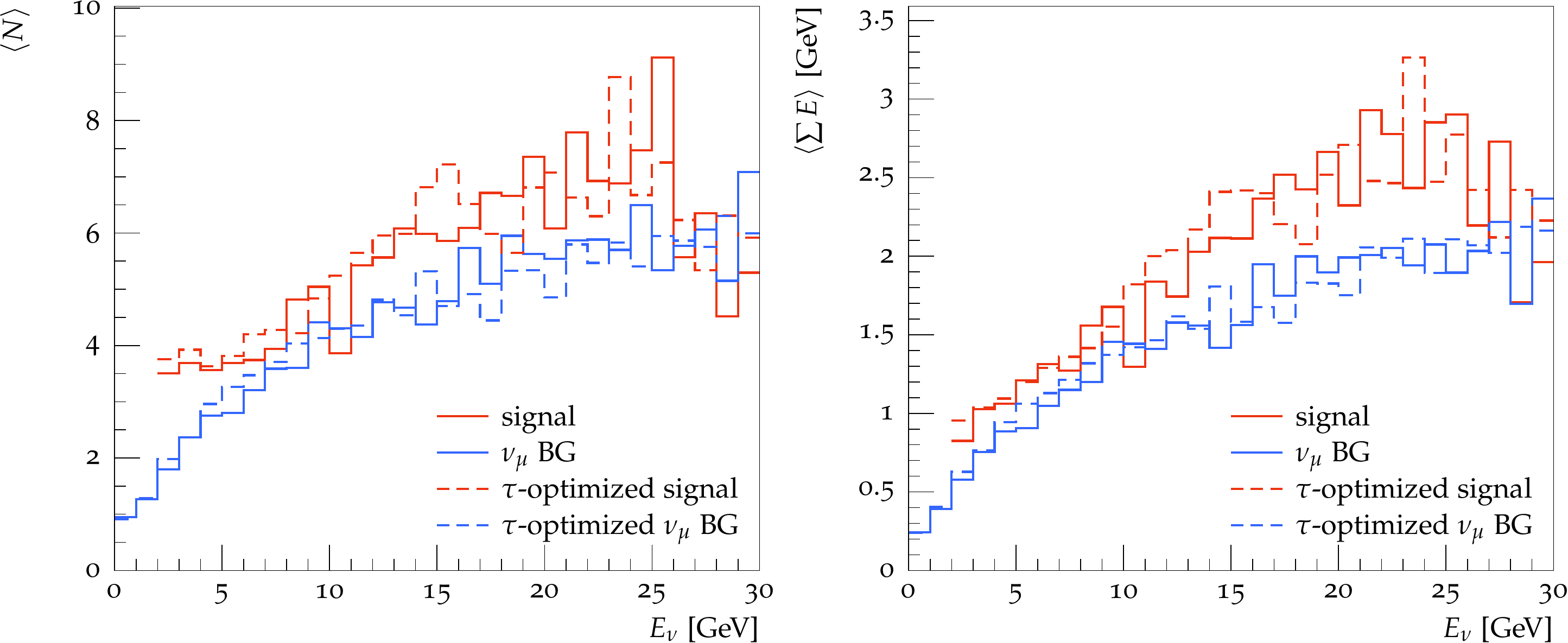}
\caption{The left (right) plot shows the  the mean number  (average sum of energies) of visible particles as a function of the true neutrino energy $E_{\nu}$ for the hadronic
channel. The signal from the nominal (tau optimized) beam mode is shown in solid (dashed) red and its dominant background  in solid (dashed) blue.
 \label{fig:profile}}
\end{figure*}
\section{Simulation and results}\label{sec:results}
In order to  account for nuclear physics effects, the signal and background neutrino-nucleon interactions are simulated
 using the GiBUU event generator~\cite{Buss:2011mx}. 
As discussed before, the signal process is a  CC interaction of a tau neutrino or antineutrino with the argon nucleus. 
This interaction produces a  tau which then decays far outside the argon nucleus. 
The Monte-Carlo events output by GiBUU  factorize into the stable, polarized tau and other final state particles  such as pions, protons and neutrons. 
The latter are products of the propagation or recoiled/created nucleons throughout the nuclear medium and are thus subject to the nuclear
 potential, re-scattering and absorption processes.

The tau lepton produced by a neutrino CC interaction will be polarized and the  distributions of its decay products will  critically depend on its spin polarization. 
Therefore it is important to consider the spin polarization of taus in addition to their production cross sections. 
This has been discussed at length 
for tau neutrinos at OPERA~\cite{Hagiwara:2003di}  and also atmospheric neutrinos at Super-Kamiokande~\cite{Aoki:2005wb} and DUNE~\cite{Conrad:2010mh}. 
We  use the TAUOLA package~\cite{Jadach:1993hs}, which decays the tau according
 to its branching ratios (see \tabref{tab:branchings}) and 
accounts for tau polarization effects.

We performed our analysis using the Rivet toolkit~\cite{Bierlich:2019rhm} which is a widely used analysis code for the LHC and other high energy physics collider experiments.  
However, we find its utility equally applicable to neutrino experiments and in this work
we premier its use at DUNE.  
Tau neutrino interactions typically lead to a high multiplicity of particles in the final state.
This is especially the case for hadronically decaying taus which exhibit significant  branching fractions to multiple mesons.
In our analysis, we apply a jet clustering algorithm to all visible final state particles. 
These include protons, charged pions, photons (as they lead to electromagnetic showers) and charged kaons but not neutrons or neutrinos. 
Jet clustering algorithms are an essential tool for a variety of LHC studies however we demonstrate
their utility  for both the $\tau_{\rm had}$ and $\tau_{e}$ channels.
We undertake this treatment for two reasons: first, the  physics of individual meson formation depends on the hadronization process, which is largely incalculable, while
 jets are objects constructed to capture the underlying hard physics which is much better understood. Second,
  applying a jet clustering algorithm via FastJet \cite{Cacciari:2011ma} is 
straightforward in Rivet and we demonstrate this technique to be a useful method of characterizing
event topologies at LArTPCs.

We divide our analysis into the hadronic and  leptonic channels and present them in \secref{sec:hadronic}
and \secref{sec:leptonic} respectively. We have used $CP$-optimized fluxes
 derived for the forward horn current polarity (neutrino mode) unless otherwise specified. 
We will also present DUNE's sensitivity to tau neutrinos in the tau-optimized beam configuration.

\subsection{Hadronic Channel}\label{sec:hadronic}
As discussed, the dominant background to the hadronically decaying taus consists of NC neutrino-argon interactions which receive
 contributions from all three neutrino flavors. 
 There are contributions from the NC interactions coming from wrong-sign neutrino contaminants
but these are subdominant as  $\overline{\nu}_e$, $\overline{\nu}_\mu$ and $\overline{\nu}_\tau$ 
 comprise $0.5\%$, $3.5\%$ and $3.3\%$ of 
 the nominal neutrino beam, respectively, in the far detector. 
 The $\overline{\nu}_\tau$ CC interactions also provide a small contribution
to the signal and this is  included in the  analysis of the hadronic channel. 

 The first step in constructing the analysis for $\tau_{\rm had}$ is to veto final state particles 
 below the following minimum energy thresholds  \cite{Abi:2020wmh}:
\begin{itemize}
 \item $\pi^{\pm}$: $E>100$ MeV
  \item $p$:  $E>50$ MeV
  \item $\gamma$, $e$, $\mu$:  $E>30$ MeV
\end{itemize}
The Monte-Carlo events contain many neutral pions, which have a decay length $c\tau\approx\,46.5\,\mu$m.
As the decay length is much larger than typical nuclear radii, GiBUU propagates the neutral pions out of the nucleus and does not decay them.
On the other hand, this decay length is too small to be resolved by DUNE and thus the $\pi^0$s decay promptly in the detector.
We decay the $\pi^0$s 
by boosting them  to their rest frame  then decaying them isotropically to two photons and  boosting the system back to the lab frame.
 \begin{figure*}[tbh]
\centering
\includegraphics[width=0.98\textwidth]{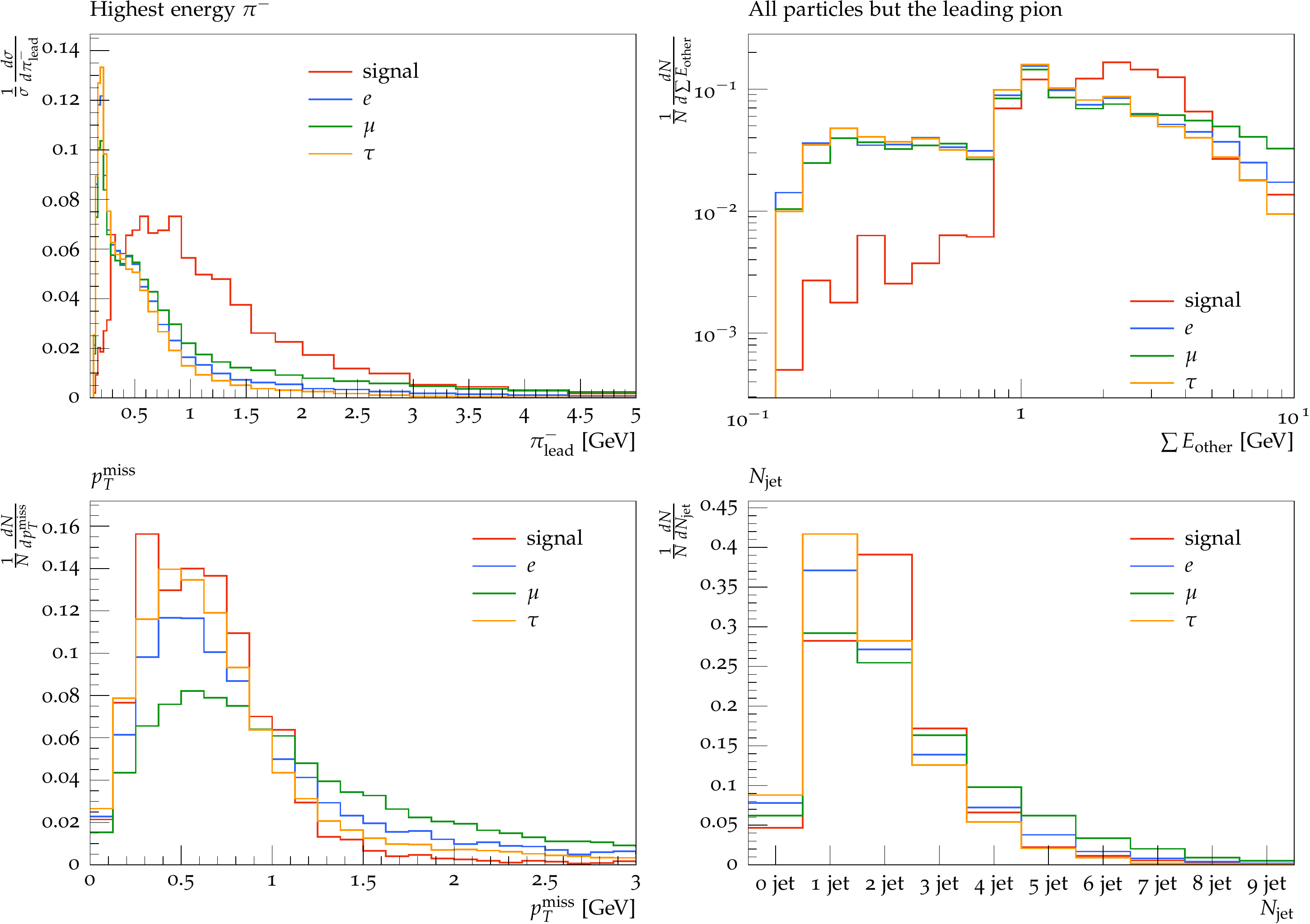}\label{fig:rivet}
\caption{The normalized distributions of the hadronic signal for the nominal neutrino flux (define as the CC $\nu_\tau$ and $\overline{\nu}_\tau$ contribution) and 
$\nu_e, \overline{\nu}_e$, $\nu_\mu, \overline{\nu}_\mu$ and $\nu_\tau, \overline{\nu}_\tau$ NC backgrounds as shown in red, blue, green and orange respectively.
In these histograms pion charge identification is assumed.}
\end{figure*}
For the $\tau_{\rm had}$ channel we have analyzed the Monte-Carlo output (as generated by GiBUU and TAUOLA) 
in terms of average multiplicity and energy sum as shown in  \figref{fig:profile}.  The left (right) plot shows  the average multiplicity (energy sum) of visible final state particles as a function of the true neutrino energy.
The events which fill these histograms have had the aforementioned thresholds applied to them. The energy threshold for tau production is evident from the 
signal (in both beam modes) which activates around $E_\nu\sim 3$ GeV.  Moreover the initial multiplicity, at low values of $E_{\nu}$, of the signal 
is $\sim 4$ which corresponds to a few visible hadrons emerging from the nuclear cascade combined with the tau dominantly decaying to two visible pions.
Unsurprisingly, the multiplicity of both the signal and background grows as a function of the true neutrino energy and is similar at high neutrino energies.
 From the right plot of \figref{fig:profile} we observe that the average visible energy sum of the signal
displays a threshold while the (dominant) background can produce low energy and multiplicity final states.  We note that at high values of the
true neutrino energy (12-25 GeV) the signal has a larger average visible energy sum than the background.
For this regime, 
in the case of the signal,  much of the true neutrino  energy will be deposited in the visible final state produced from the tau decays. 
However, in the case of the background, 
for the same value of true neutrino energy, all the deposited energy will 
result in the hadronic shower which can contain invisible neutrons. 

In order to optimize the signal-to-background ratio, we study the distributions of several kinematic variables
initially assuming charge identification of the pion is possible. 
Here we provide a list of  these variables and the cuts we apply:
\begin{enumerate}
  \item \underline{$N_{\rm lep}$} is the number of $e^\pm$ and $\mu^\pm$.  We veto events containing any such leptons in the final state.
  \item \underline{$N_{\pi^-}$} is the number of $\pi^-$'s in the final state. We veto events containing zero $\pi^-$.
    \item \underline{$\pi^-_{\rm lead}$} is the energy of the leading (highest energy) $\pi^-$ in each event. We veto events if the leading $\pi^-$ has $E<250$ MeV.
     \item \underline{$\sum E_{\rm other}$} is the total visible energy of the event \emph{excluding} the leading $\pi^-$. We veto events with  $\sum E_{\rm other}< 600$ MeV.
     \item \underline{$p^{\rm miss}_{T}$} is the  missing transverse momentum. We veto events with $p^{\rm miss}_{T} > 1$ GeV.
     \item  \underline{$N_{\rm jet}$} is the number of jets in the final state. We veto zero jet events. 
 \end{enumerate}

   We note that the cuts are applied in this order.  We also consider the possibility that charge identification 
 is not possible and alter the analysis  such that $\pi^-$ is replaced with $\pi^\pm$.

 For the signal, we simulated the tau decaying to all possible final states.
  Therefore, approximately $35\%$ of the events contain electrons and muons and the first cut removes this leptonic contribution. To construct 
 the remaining cuts we considered that the dominant tau decays (constituting $\sim 70\%$ of the hadronic channel)
  are $\tau^-\to \pi^-\nu_\tau$, $\tau^-\to \pi^-\pi^0\nu_\tau$ and $\tau^-\to \pi^-\pi^0\pi^0\nu_\tau$ as shown in
   \tabref{tab:branchings} and the signal tends to contain a hard $\pi^-$ which motivates the  second  and third cuts.
 The  normalized distributions of the kinematic variables   of the signal (red) and electron,
  muon and tau (blue, green and orange respectively) backgrounds are shown in \figref{fig:rivet}. We note that
   cuts have not yet been applied to the 
 events that fill these histograms other than  $(1)$ and $(2)$. These distributions show vital shape information that we used to design cuts (3)-(6).
   The upper left plot shows the normalized distributions of energies of the leading $\pi^-$ and it is clear
that the signal has a larger proportion of high energy $\pi^-$ while
 the background has a distinct peak  in the low energy bins. For the signal, the hardest $\pi^-$
 originates from the tau decay and it carries $E\gtrsim \mathcal{O}(100)$ MeV in energy while the background
 is characterized by many lower energy hadrons resulting from the 
  hadronization process followed by propagation of these hadrons through the nucleus via intra-nuclear cascade.  A crossover between the signal and background distributions 
 occurs around  $E_{\pi_{\rm lead}^-}\sim250$ MeV and we place our cut here to enrich the signal and deplete the background. 

The second observable we consider is $\sum E_{\rm other}$ and its corresponding  normalized distribution is shown in
the upper right plot of \figref{fig:rivet}. The shape difference between the signal and background is distinct:  the background is 
relatively flat in this observable apart from a slight increase in the distribution around $1$ GeV. 
In contrast, the signal distribution has a marked dip below $1$ GeV. 
We can observe this dip derives
from the tau production threshold as shown in the right plot of \figref{fig:profile} where the average visible energy 
has an initial value $\sim 1$ GeV for true neutrino energies close to the tau production threshold.
For $\sum E_{\rm other}>1$ GeV we observe the signal increases and this
 corresponds to the difference in the signal and background spectra as shown in the right plot of \figref{fig:profile}. We note that the reason
for this difference is the same in this observable as it was in the aforementioned histogram. 
We varied the cut in this observable between $200 \leq\sum E_{\rm other} (\text{MeV}) \leq 800$ and found  that a veto on events with $\sum E_{\rm other} < 600$ MeV reduces the backgrounds most effectively.
  
  The distribution of the  missing transverse momentum is shown in the lower left plot of \figref{fig:rivet}.
The invisible states which contribute to this missing transverse momentum vector are neutrinos and neutrons~\footnote{Fermi momentum $\mathcal{O}(200~{\rm MeV})$ in the argon nucleus may also contribute to the missing transverse momentum.}. 
We observe that the 
  signal distribution is more strongly peaked than the background as the former will always have a $\nu_\tau$ in the final state. 
  A crossover in the shapes of the signal and background occurs at $p^{\rm miss}_{T}\sim 900$ MeV which motivates cut (5). 

  The final cut is on the jet multiplicity and we discuss  this in more detail due to its non-standard application in neutrino physics.
 Rivet uses FastJet to cluster visible final states  into  jets. We define the jet to   have a minimum energy of $1$ GeV. In particular, 
 we use the Cambridge-Aachen algorithm \cite{Dokshitzer:1997in} which falls into the $k_{T}$ class of jet clustering algorithms.  
In order to be clustered into a single jet, visible particles must be within a  radius of $R=0.6$, with $R=\sqrt{\eta^2+\phi^2}$ where $\eta=-\log\tan(\theta/2)$ is the pseudo-rapidity ($\theta$ is the angle between the particle and the jet axis) and $\phi$ is
the azimuthal angle with respect to the jet axis~\footnote{We varied the radius in the interval $0.2\leq R\leq1.0$ and
 found the analysis was not particularly sensitive to the jet size in this range.}.  The lower right plot of \figref{fig:rivet} shows the
  distribution of  jet multiplicities. We observe that the signal  has a lower zero jet rate than any of the $e$, $\mu$ or $\tau$ backgrounds.
   Moreover, the background peaks  at one jet while the signal peaks at two jets. We can interpret this in the following way: 
    the background predominantly has a single jet
which emerges from the argon nucleus (see the lower left image of \figref{fig:diagrams} where this jet is denoted as $n_{\rm jet}$). 
On the other hand, the signal mainly has a two-jet final state where one jet is produced from the  tau decay (denoted as $\tau_{\rm jet}$ in the upper left image of \figref{fig:diagrams}) and  
another from the intranuclear 
cascade. The number of jets in the final state can be as high as nine, although
these higher jet multiplicities are  suppressed. This is due to the broad span of the neutrino beam energy. 
 We found the optimal cut, for this definition of a jet, is to veto zero jet events. 
   \begin{figure}[t!]
\centering
\includegraphics[width=0.48\textwidth]{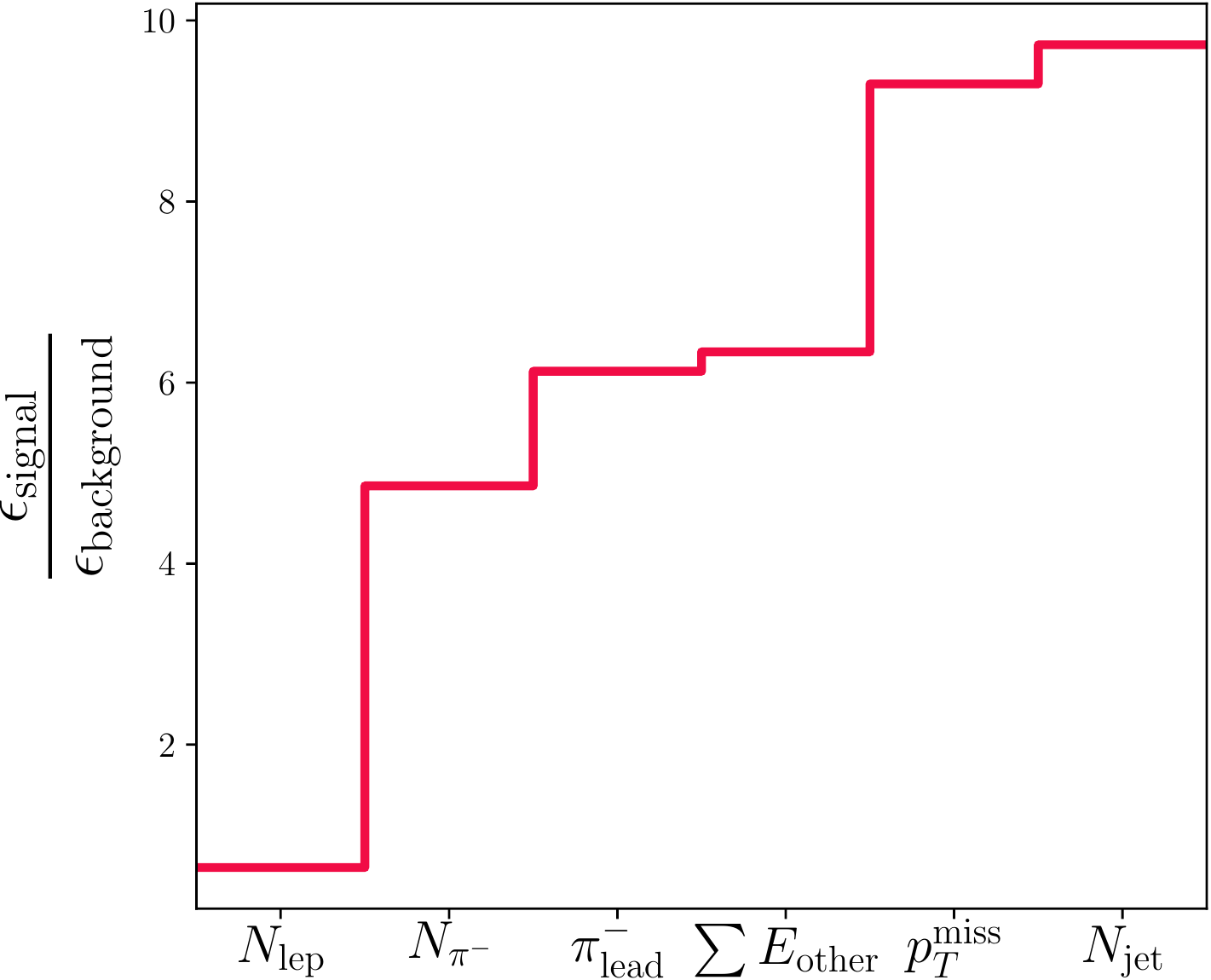}\label{fig:C&C}
\caption{Cut flow analysis for the hadronic signal and associated NC backgrounds (using the nominal neutrino flux) assuming charge identification of the pion.}
\end{figure}

The efficacy of the cuts is summarized in \figref{fig:C&C}.
We show the ratio of  efficiencies of the signal over the NC background interactions, as a function of each cut.
The application of the cuts to the events should be read from left  to right. At this stage,
fluxes are not taken into account and this plot simply represents the effectiveness of each cut. 
The first cut has a value of $\epsilon_{\rm signal}/\epsilon_{\rm background}\sim 0.65$ which results
from  removing the 
leptons from the signal sample. The second cut, which ensures there is at least one negatively charged pion in the final state, 
drastically increases the efficiency from $\sim 0.65$ to $\sim 4.8$ and we find this to be the most aggressive requirement on final states. 
The cut on the energy of the leading $\pi^-$
and high values of missing transverse momenta also have a significant effect. Further, we find the requirement of at least a single jet in the 
final state is more effective than vetoing $\sum E_{\rm other}< 600$ MeV.  

 \begin{figure}[t!]
\centering
\includegraphics[width=0.48\textwidth]{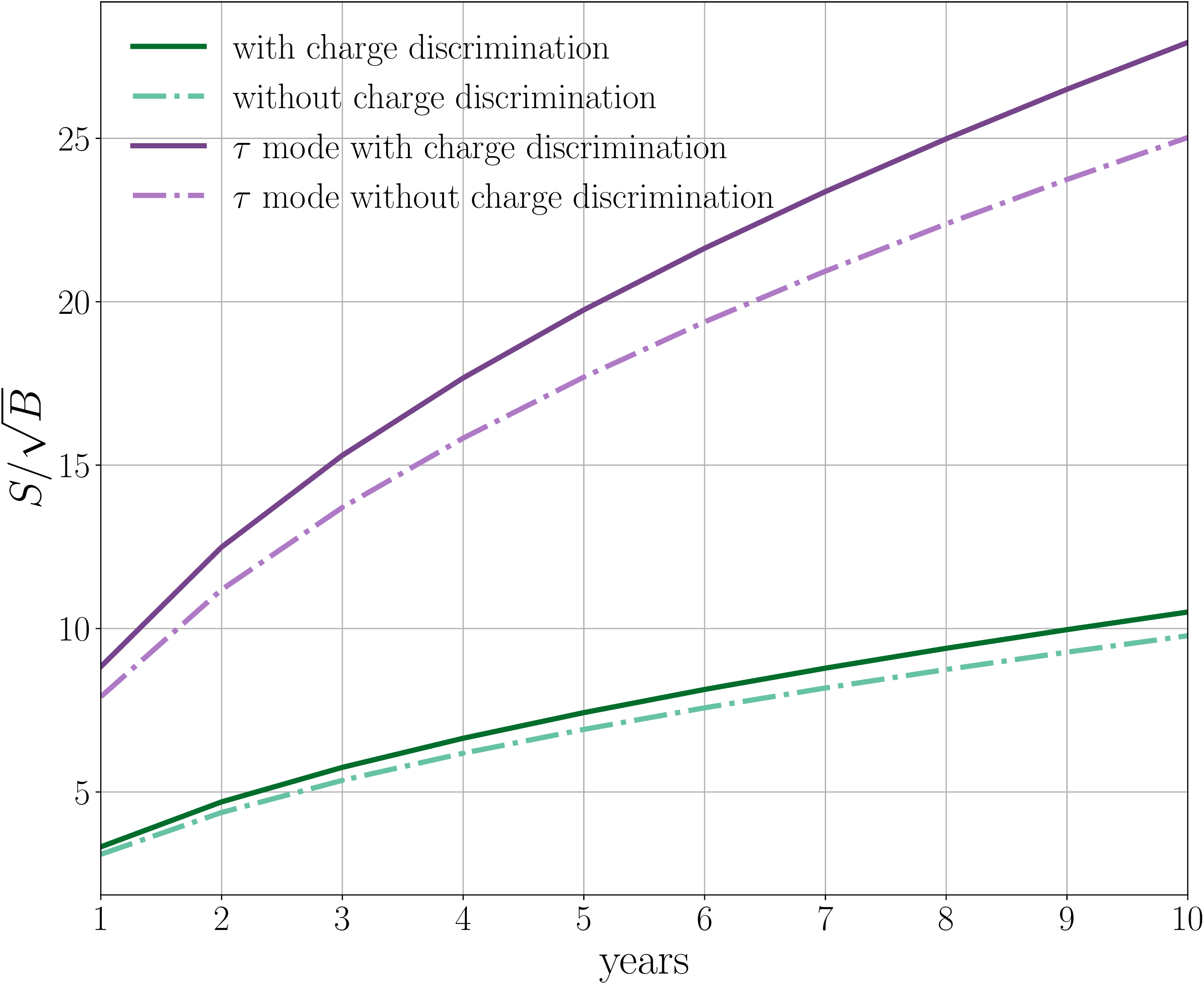}
\caption{The solid green (solid purple) shows the signal over square root of background as a function of time for the nominal neutrino (tau optimized) mode assuming pion charge identification.
The dot-dashed green (dot-dashed purple) shows the signal over square root of background as a function of time for the nominal neutrino (tau optimized) mode without pion charge identification. \label{fig:SoverBHad}}
\end{figure}

Finally, to demonstrate the impact of these series of cuts on $\nu_\tau$/$\overline{\nu}_\tau$ detection at DUNE, we show the
significance, defined as $S/\sqrt{B}$ where $S$ ($B$) is the number of signal (background) events,  as a function of run time in years in \figref{fig:SoverBHad}. 
The solid dark  green shows the significance
 for the nominal beam configuration if DUNE has the capability to distinguish $\pi^-$ from $\pi^+$  and  we find  $S/\sqrt{B}\sim 3.3$ within a year of data-taking. 
This corresponds to the detection of $79$  and $565$ signal and background events respectively.
In the scenario that charge identification is not possible, as indicated by dashed light green,  the number of signal and background events detected after one year is $83$ and  $731$ respectively.
Therefore, the significance  decreases to  $S/\sqrt{B}\sim 3.1$ and a significance value of $5$ requires approximately
$2.5$ years of data-taking. 
The improvement in the significance with charge identification  is 
 mainly due to background reduction, as multi-GeV neutral current interactions will tend to produce comparable amounts of leading $\pi^+$ and $\pi^-$, while the signal is dominated by leading $\pi^-$.
Thus, pion charge identification can be used to further mitigate backgrounds without reducing the signal.
In the optimal scenario, a perfect pion charge discrimination would be equivalent to an increase of about $17\%$ in exposure in tau neutrino analyses.
This result demonstrates the non-trivial leverages that LArTPCs may have when making full use of topological information.

We applied the same analysis cuts to the
tau-optimized beam sample and found the significance to be almost three times higher compared to the nominal beam.  The solid dark
purple shows the significance with charge discrimination and 
after one year of data-taking is $\sim 8.8$. This corresponds to $433$ and  $2411$ signal and background events respectively.  Unsurprisingly, without the charge discrimination capability the significance is lower with a value of $\sim7.9$ which
corresponds to $439$ and $3077$ signal and background events respectively after the first year of running in tau-optimized mode.
\subsection{Leptonic Channel}\label{sec:leptonic}
\begin{figure}[t!]
\centering
\includegraphics[width=0.48\textwidth]{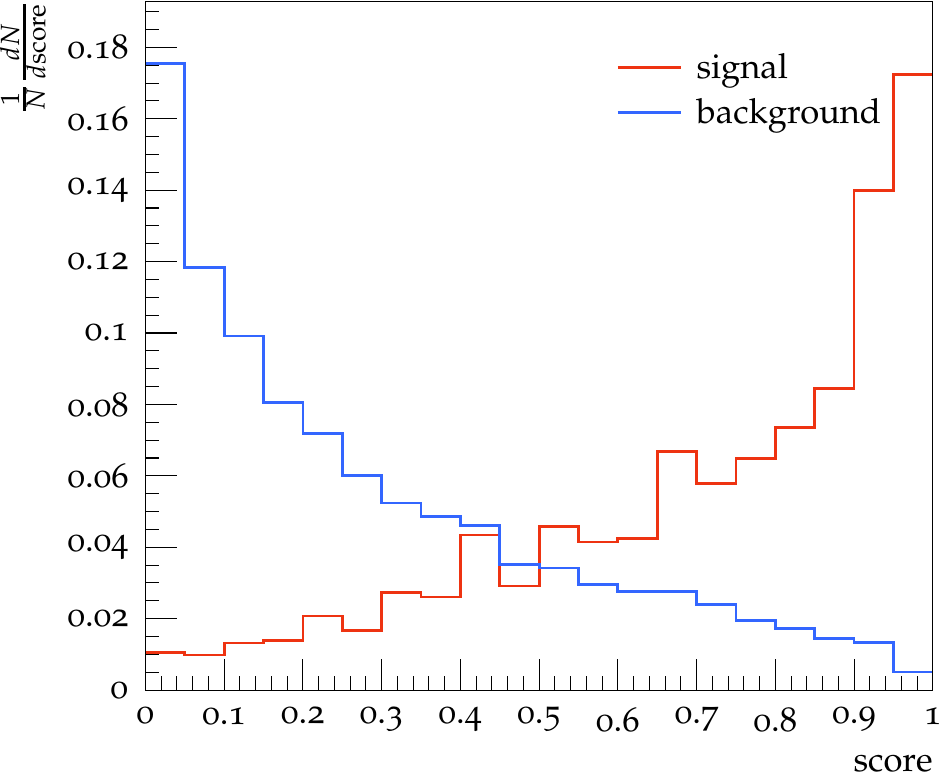}\label{fig:score_nom}
\caption{The normalized distributions of the $\tau_e$ signal (red) and associated background (blue)  for the nominal neutrino beam.}
\end{figure}

As outlined before, the 
leptonic decay channels of the tau are  more challenging than their hadronic counterparts. First,  
the  background  cross-section (CC interactions from $\nu_e$ and $\nu_\mu$)
 is  larger than the NC background.  Second,
 the tau decays to  charged leptons at approximately half the rate as it does to hadrons. 
 Nonetheless, we pursue the $\tau_e$ channel where the dominant background is the CC interaction
 of electron neutrinos. We neglect the  contribution from the  $\overline{\nu}_e$ CC
events as the $\overline{\nu}_e$ composition of the neutrino beam is  approximately $0.5\%$ at the far detector.
  We attempted to construct a simple cut and count analysis for $\tau_e$, in a similar manner
 to the $\tau_{\rm had}$ analysis, but
 we  found the 
 significance after one year of data-taking was below 1.0. 
 In light of this,  a more effective way to discriminate the signal from the background is to use  a deep
 neutral net (DNN). In particular, we utilize Keras with tensorflow \cite{chollet2015keras}. Our methodology is as follows:
\begin{enumerate}
\item Generate signal and background Monte-Carlo samples. 
\item Use Rivet, with the same minimum energy thresholds as before,  to calculate kinematic variables  or ``features" of   
background and signal events. The signal is assigned a ``classification variable" value  of $1$ and the background $0$.
\item Separate data sets into training, validation and test samples. 
\item We train the DNN using the binary cross entropy loss function with the training data.  We use the validation data set to guard against overtraining.
\item We feed the test data through the trained DNN and for each event it returns a ``score" between $0$ and $1$. If the event is more
background-like its score is closer to $0$ and conversely if it is more signal-like its score is closer to $1$.
\end{enumerate}
The score can be thought of as a new observable that allows for discrimination of signal and background events.  The kinematic 
variables or \emph{features} used to characterize the signal and background are
  \begin{figure}[t!]
\centering
\includegraphics[width=0.48\textwidth]{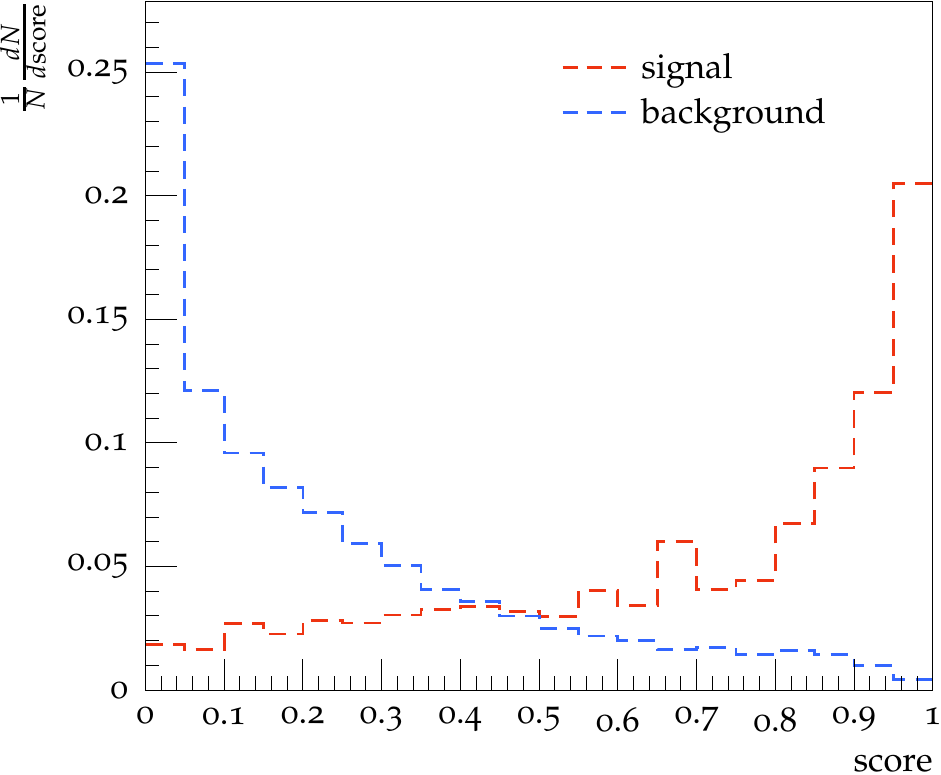}\label{fig:score_TO}
\caption{The  normalized distributions of the $\tau_e$ signal (red) and associated  background (blue) for the tau-optimized neutrino beam.}
\end{figure}
\begin{itemize}
\item \underline{$d\phi_{\rm min}$} is the minimum angle between the leading (highest energy) $e^-$ and any other visible particle. 
\item  \underline{$dR_{\rm min}$} is the $\Delta R$ between leading $e^-$ and any other visible particle.
\item  \underline{$d\phi_{\rm met}$} is the $\Delta \phi$ between leading lepton and the missing transverse momentum vector.
\item  \underline{$E^\ell_{\rm lead}$} is the leading  lepton energy. 
\item \underline{$E^{\rm miss}_{T}$} is the missing transverse energy.
\item \underline{$N$} is the number of visible particles other than leading lepton.
\item \underline{$N_{\pi^\pm}$} is the number of $\pi^\pm$'s.
\item \underline{$\sum E_{\rm other}$} is the sum of the energies of all visible particles other than leading lepton.
\item \underline{$\theta_{\ell}$}  is the angle of the lepton with respect to the beam axis.
\item \underline{$N_{\rm jet}$} is the number of jets where jets have the same definition as in the hadronic channel.
\item \underline{$E^{\rm jet}_{\rm lead}$} is the energy of the leading jet. 
\end{itemize}

We note that the lepton is not included in the jet definition and we do not apply charge identification of
the pion in this analysis as the number of negatively and positively charged pions should
be approximately the same for the $\tau_e$ signal and background. 
Our chosen DNN architecture is a sequential model with a dense
input layer followed by two
hidden layers of depth 100 sandwiching a dropout layer with a dropout parameter of value 0.2. 
The final layer is a sigmoid output layer of depth 1.

 \begin{figure}[t!]
\centering
\includegraphics[width=0.49\textwidth]{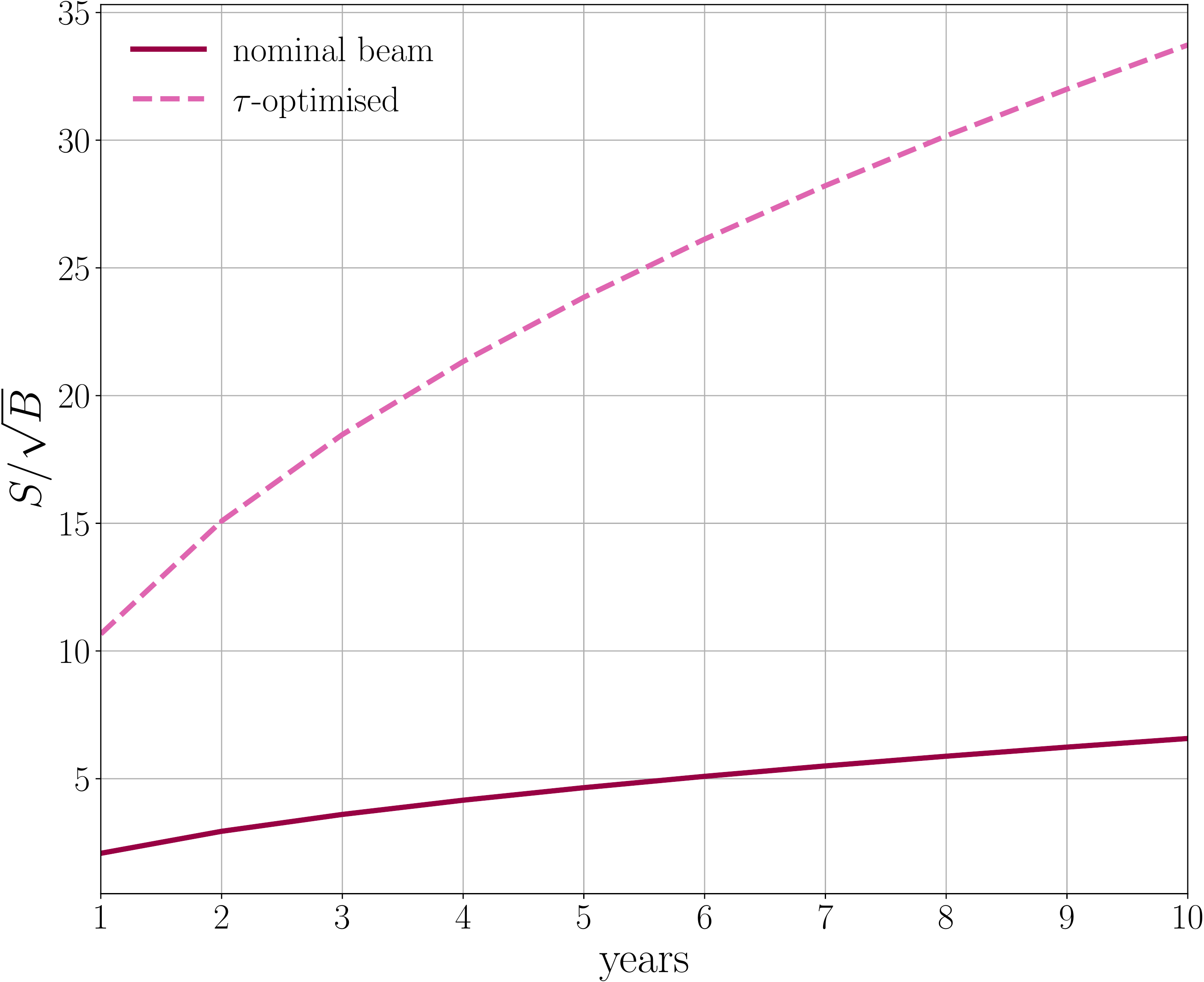}\label{fig:DNNsoverb}
\caption{The solid purple (dashed pink) shows the  signal over square root of background as a function of time for the $\tau_e$ channel using the nominal (tau-optmized) neutrino flux.}
\end{figure}

 The normalized score distributions of signal and background, for nominal neutrino beam mode, are shown in \figref{fig:score_nom}.  As expected, the score value of the background peaks at $0$ 
 and the signal peaks at $1$. 
  We varied where we placed the cut on the score observable and found vetoing events with 
 a score less than $0.85$ yields the highest significance with $S/\sqrt{B} = 2.3$ after one year of data-taking corresponding to
 $13$ signal events and $32$ background events.
 The significance as a function of time is given by the solid dark purple line of \figref{fig:DNNsoverb} where we observe a 
  $S/\sqrt{B}=5.0$ can be reached after six years of running. 

 As with the case of the hadronic channel, the detectability of tau neutrinos is vastly improved using the 
tau-optimized beam.  The normalized score distributions of signal and background are shown in \figref{fig:score_TO}. 
We found that vetoing score values below $0.8$ gave us the optimal  significance of $11.0$ corresponding to $63$ signal and 
$33$  background events, respectively, after one year of data-taking. 
The significance is indicated by the dashed pink line  of \figref{fig:DNNsoverb}. 

Compared to the hadronic tau sample, the $\tau_e$ sensitivity using the nominal beam is smaller but  may still  provide valuable additional information on tau neutrinos. It is remarkable and somewhat surprising  that using  the tau-optimized beam,
the sensitivity of the $\tau_e$ sample is comparable to the $\tau_{\rm had}$ sensitivity (comparing \figref{fig:SoverBHad} to \figref{fig:DNNsoverb}).
The significant enhancement of $S/\sqrt{B}$ for the $\tau_e$ channel in the high energy run can be understood qualitatively by comparing the hadronic and leptonic tau analyses. 
The backgrounds to $\tau_{\rm had}$ comes from NC events which are flavor blind. 
Compared to the nominal beam mode, the tau-optimized run presents a higher value of the signal-to-background ratio because there is a larger fraction of the $\nu_\tau$ flux above the tau production threshold.
This enriches the signal significantly. 
On the other hand, the $\tau_e$ channel receives background contribution mainly from CC $\nu_e$ events.
This background is strongly affected by oscillations, as the $\nu_e$ contamination in the initial neutrino beam is very small.
At high energies, $\nu_\mu\to\nu_e$ oscillations are suppressed by $1/E_\nu^2$, see  \equaref{eq:oscillations}. 
Therefore, operating in the tau-optimized mode in the $\tau_e$ case not only significantly enriches the signal, but also strongly depletes the background.
This effect can be appreciated in \figref{fig:FDflux} by comparing the $\nu_e$ (blue) and $\nu_\tau$ (red) fluxes at DUNE far detector for the nominal (solid) and tau-optimized (dashed) runs.
  \begin{table}[t]
\begin{center}
\begin{tabular}{|c|c|c|c|c|c|} \hline
   \textbf{Mode} & \textbf{beam}& \textbf{charge id}  &  $N_{\rm sig}$ &$N_{\rm bg}$ & $S/\sqrt{B}$ \\ \hline \hline
 $\tau_{\rm had}$ & nominal & \cmark & 79 &565 &  3.3 \\
  $\tau_{\rm had}$ & nominal & \xmark & 83 &731 & 3.1 \\
      $\tau_{\rm had}$ & tau-optimized   & \cmark &433 &2411 & 8.8\\
    $\tau_{\rm had}$ & tau-optimized   & \xmark & 439 &3077& 7.9 \\
    \hline \hline
     $\tau_{e}$ & tau-optimized  & \xmark & 63 & 33 & 11.0\\
      $\tau_{e}$ & nominal   &\xmark  & 13 &32 &  2.3 \\
      \hline
  \end{tabular}
\end{center}
\caption{The number of signal and background events after one year of data-taking in the hadronic ($\tau_{\rm had}$) and leptonic ($\tau_{e}$) channels
for nominal and tau-optimized beam modes.\label{tab:summaryresults}}
\end{table}

\section{Discussion and Future Prospects}\label{sec:summary}
The analysis strategy proposed in this paper demonstrates the vast potential of  LArTPC capabilities.
There are two clear avenues for further exploration.
First, although we performed an in-depth \emph{physics} analysis of tau neutrinos at DUNE, a realistic simulation of detector effects is required to completely demarcate DUNE's sensitivity to beam $\nu_\tau$ appearance.
We intend to explore, in a forthcoming work, the impact of detector effects on the kinematic variables used here and consequently the signal-to-background ratio.
Second, in order to perform a more robust physics analysis with the tool chain we developed here, a detailed understanding of tau neutrino energy reconstruction is needed 
in addition to the impact of systematic uncertainties on the inferred  $\nu_\tau$ spectrum.

Special attention should be given to $\mu^\pm/\pi^\pm$ mis-identification which can  enlarge the $\tau_{\rm had}$ background.
Moreover,   $e/\gamma$ separation is needed to reject certain NC events as a background in the $\tau_e$ analysis.
This analysis, excluding  the missing transverse momentum information, could also be  implemented  in multi-GeV atmospheric neutrino searches in DUNE and Hyper-Kamiokande~\cite{Abe:2018uyc}, with some tuning of the analysis cuts due to much higher incoming neutrino energy.

Ultimately, several physics analyses could benefit from applying our tool chain with realistic detector effects and a treatment of systematic uncertainties.
Amongst those are $\nu_\tau$-nucleus interaction measurements~\cite{Jeong:2010nt, Sobczyk:2019urm}; violations of unitarity in the leptonic mixing matrix~\cite{Farzan:2002ct, He:2013rba, Qian:2013ora, Parke:2015goa, Ellis:2020ehi}; non-standard neutrinos interactions in neutrino production, detection and propagation~\cite{Dev:2019anc}; sterile neutrino searches~\cite{Collin:2016rao, Gariazzo:2017fdh, Dentler:2018sju, Ghoshal:2019pab}; and general consistency tests of the three neutrino oscillation paradigm~\cite{deGouvea:2019ozk, Denton:2020exu}.
In our work we demonstrate that the tau-optimized beam significantly enhances the prospects of tau neutrino measurements at DUNE. In particular, running in 
this mode improves the detectability of tau neutrinos in the leptonic channel.
A successful case of a high energy neutrino beam run can be found in the MINOS experiment history.
For  example, the constraints set by MINOS+, the high energy run of MINOS, on eV-scale sterile neutrinos~\cite{Adamson:2017uda} and large extra dimension models~\cite{Adamson:2016yvy} remain leading the field even in the presence of newer experiments.

\section{Conclusions}
We have proposed a novel analysis strategy to study tau neutrinos with the DUNE experiment.
The marriage of collider tools and neutrino physics allows us to exploit topological features in neutrino events and 
thereby significantly extend DUNE's physics reach. The Rivet analysis code 
used for the hadronic channel is available at \cite{holger_schulz_2020_3924078}.
Our results are summarized in \tabref{tab:summaryresults} and we find that, in the nominal neutrino beam, DUNE could achieve $S/\sqrt{B}=5$ in just over two years of data-taking using only hadronically decaying taus or in six years of data-taking with leptonically decaying  taus.
In the tau-optimized beam mode, after a single year of data-taking, DUNE could achieve $S/\sqrt{B}\sim 9, 11$  for the $\tau_{\rm had}$ and $\tau_e$ channels respectively.
Moreover, identifying the pion charge via pion track topology in DUNE would be equivalent to an increase between 17\% and 24\% in exposure. 
\acknowledgements

We  thank  Kai Gallmeister, Ulrich Mosel and Zbigniew Andrzej Was  for their invaluable advice on the usage of 
GiBUU and TAUOLA.  We are grateful to  Andr\'{e} de Gouv\^{e}a for careful reading of the manuscript. We thank  Kirsty Duffy,  Lukas Heinrich, Joshua Isaacson, Xiangyang Ju, Kevin Kelly, Claudius Krause, Shirley Li, Nhan Tranh, Raquel Castillo Fernandez and Tingjun Yang for helpful discussions on various aspects of this work.

This research was supported by the Fermi National Accelerator Laboratory 
(Fermilab), a U.S. Department of Energy, Office of Science, HEP User Facility.
Fermilab is managed by Fermi Research Alliance, LLC (FRA), acting under 
Contract No. DE--AC02--07CH11359. This material is based upon work supported by the U.S. Department of Energy, Office of Science, Office of Advanced Scientific
Computing Research, Scientific Discovery through Advanced Computing (SciDAC) program, grant ÒHEP Data Analytics on HPCÓ,
No. 1013935. It was supported by the U.S. Department of Energy
under contracts DE-AC02-76SF00515.

\bibliographystyle{apsrev4-1}
\bibliography{ref}{}
\end{document}